\documentclass[aip, apl, reprint]{revtex4-1}

\usepackage{graphicx}
\usepackage{subfigure}
\usepackage{hyperref}
\usepackage{siunitx}
\usepackage{xcolor}
\usepackage[utf8]{inputenc}
\usepackage[T1]{fontenc}
\usepackage{ulem}
\usepackage{amsmath}

\graphicspath{{../Figures/PDF/}}
\makeatletter
\let\@fnsymbol\@fnsymbol@latex
\@booleanfalse\altaffilletter@sw
\makeatother

\begin{document}
\sisetup{range-phrase=--}

\title{Direct loading of nanoparticles under high vacuum into a Paul trap for levitodynamical experiments} 



\author{Dmitry S. Bykov}
\email{dmitry.bykov@uibk.ac.at.}
\affiliation{Institut für Experimentalphysik, Universität Innsbruck, Technikerstraße 25, 6020 Innsbruck, Austria}
\author{Pau Mestres}
\affiliation{Institut für Experimentalphysik, Universität Innsbruck, Technikerstraße 25, 6020 Innsbruck, Austria}
\author{Lorenzo Dania}
\affiliation{Institut für Experimentalphysik, Universität Innsbruck, Technikerstraße 25, 6020 Innsbruck, Austria}
\author{Lisa Schmöger}
\affiliation{Institut für Experimentalphysik, Universität Innsbruck, Technikerstraße 25, 6020 Innsbruck, Austria}
\affiliation{Max-Planck-Institut für Kernphysik, Saupfercheckweg 1, 69117 Heidelberg, Germany}
\author{Tracy E. Northup}
\affiliation{Institut für Experimentalphysik, Universität Innsbruck, Technikerstraße 25, 6020 Innsbruck, Austria}

\date{\today}

\begin{abstract}
Mechanical oscillators based on levitated particles are promising candidates for sensitive detectors and platforms for testing fundamental physics. The targeted quality factors for such oscillators correspond to extremely low damping rates of the center-of-mass motion, which can only be obtained if the particles are trapped in ultra-high vacuum (UHV). In order to reach such low pressures, a non-contaminating method of loading particles in a UHV environment is necessary. However, loading particle traps at pressures below the viscous flow regime is challenging due to the conservative nature of trapping forces and reduced gas damping. We demonstrate a technique that allows us to overcome these limitations and load particles into a Paul trap at pressures as low as \SI{4e-7}{\milli\bar}. The method is based on laser-induced acoustic desorption of nanoparticles from a metallic foil and temporal control of the Paul trap potential. We show that the method is highly efficient: More than half of trapping attempts are successful. Moreover, since trapping attempts can be as short as a few milliseconds, the technique provides high throughput of loaded particles. Finally, the efficiency of the method does not depend on pressure, indicating that the method should be extensible to UHV.
\end{abstract}


\maketitle 
Recently, building on early experiments of Ashkin with optically suspended particles~\cite{Ashkin1970}, a new research direction of levitated optomechanics or levitodynamics has emerged~\cite{Neukirch2015,Bhattacharya2017}, for which a central challenge is to bring the motion of levitated particles into the quantum realm~\cite{ChangRegalPappEtAl2010,Romero-Isart2010,Barker2010,Romero-Isart2011a}. Towards this goal, the techniques employed for particle trapping have been extended beyond optical levitation to include methods such as magneto-gravitational traps~\cite{Lyuksyutov2004,Hsu2016} and Paul traps~\cite{Kane2010,KuhlickeSchellZollEtAl2014,MillenFonsecaMavrogordatosEtAl2015,Delord2017b,Conangla2018}. Levitated particles are unique among mechanical oscillators in that they benefit from the absence of physical contact with the environment, which provides an opportunity to achieve high quality factors at room temperature if a UHV \SIrange[range-units = single]{e-9}{e-12}{\milli\bar} environment is used~\cite{ChangRegalPappEtAl2010,Romero-Isart2010}. For example, background-gas-limited Q factors can be as high as $10^{12}$ if pressures of \SI{e-10}{\milli\bar} are achieved~\cite{KieselBlaserDelicEtAl2013}. Moreover, these pressures are similar to those required for laser cooling and trapping of atoms and ions; particle trapping at UHV thus opens up opportunities for hybrid systems of mechanical oscillators coupled to atoms~\cite{Wallquist2009,Genes2009,Pflanzer2013}.

Regardless of the trapping mechanism, such low pressures impose constraints on the particle loading method. Several techniques are currently used to load particles into optical, magnetic, or Paul traps, including spraying solutions of the particles with a medical nebulizer~\cite{SummersBurnhamMcGloin2008} or an electrospray~\cite{Y.Cai2002}, launching the particles from surfaces using piezoelectric transducers~\cite{AshkinDziedzic1971} or laser-induced acoustic desorption (LIAD)~\cite{AsenbaumKuhnNimmrichterEtAl2013,Millen2016}, and transferring pre-trapped particles to a main trap by means of load-lock techniques~\cite{MestresBerthelotSpasenovicEtAl2015} or hollow-core fibers~\cite{benabid_particle_2002, BykovSchmidtEuserEtAl2015, GrassFeselHoferEtAl2016}. However, all of these methods have disadvantages when applied in a UHV-compatible setup. In the case of the first two techniques, spraying solvents into a vacuum chamber will introduce contaminants that are difficult to pump, e.g., water molecules. Moreover, due to the conservative nature of gradient trapping forces, particle capture requires a dissipation mechanism. This dissipation is typically provided by a background gas; as a result, the solvents have to be sprayed at ambient pressure or in low vacuum. Operation at UHV would thus require cycling the vacuum setup between low vacuum and UHV, a time-consuming procedure that introduces additional experimental complexity. Launching from surfaces using piezoelectric transducers and LIAD provides a dry loading method. However, both of these methods still require low vacuum or ambient pressure for initial trapping. The load-lock techniques provide a possibility to transfer particles from a contaminated chamber to a UHV-compatible one, but with the drawback of a bulky setup with limited throughput at low pressures. Hollow-core photonic crystal fibers offer a compact, flexible solution with throughput better than that of load-lock techniques. However, stable particle trapping in such fibers at pressures below $\SI{e-2}{\milli\bar}$ has yet to be demonstrated.

\begin{figure*}
 \includegraphics[width=1.8\columnwidth]{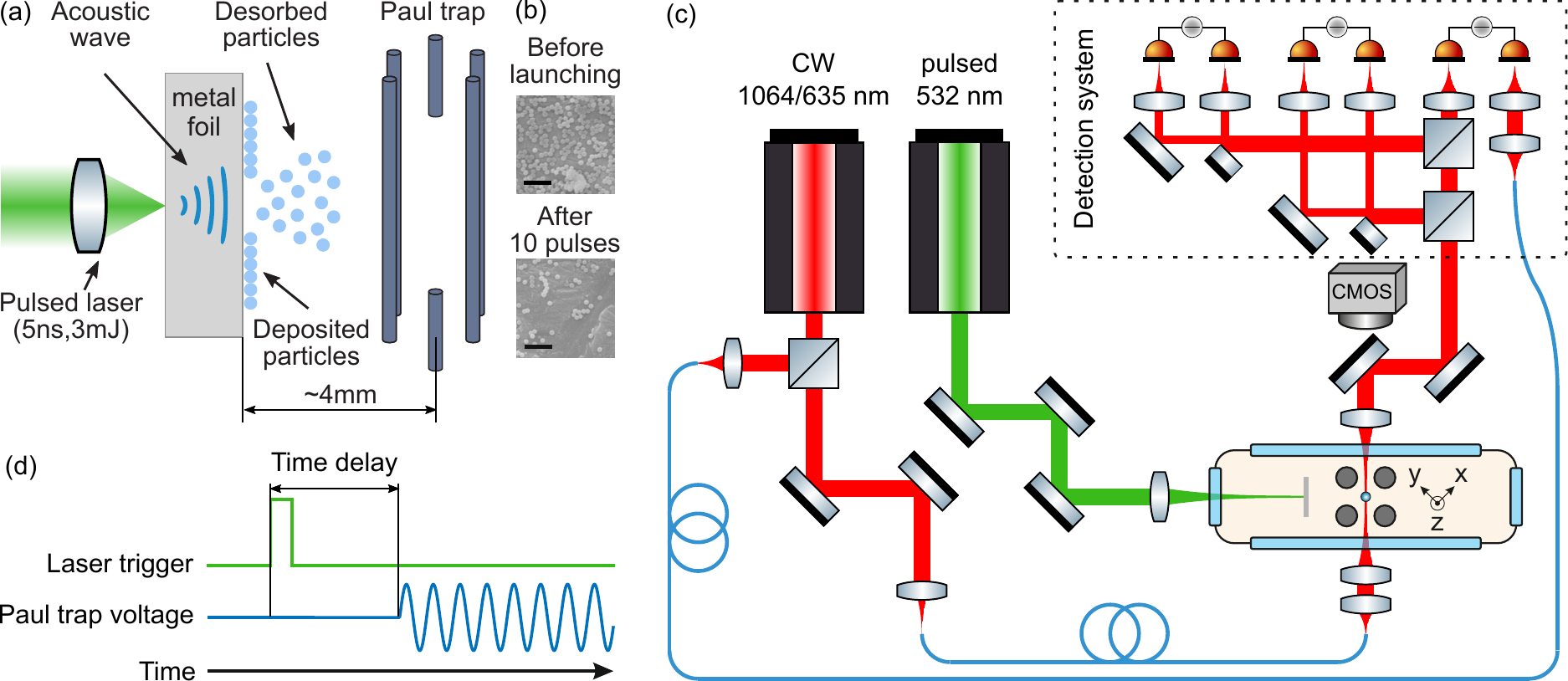}
 \caption{Loading procedure and experimental setup. (a) Schematic of the laser-induced acoustic desorption (LIAD) procedure, in which deposited particles are ejected from a thin aluminum foil into the ion trap region. (b) Top: Scanning-electron-microscope (SEM) image of foil on which $\SI{300}{\nano\meter}$ diameter nanoparticles have been deposited. Bottom: the same foil after execution of ten LIAD pulses. The scale bar is \SI{2}{\micro\meter}. (c) Experimental setup.  A pulsed \SI{532}{\nano\meter} laser is focused on the back side of the foil to desorb particles. A continuous wave (CW) \SI{635}{\nano\meter} laser illuminates the particles, which are imaged by a CMOS camera. The motion of trapped particles is monitored using a focused CW \SI{1064}{\nano\meter} laser beam and a detection system based on back-focal-plane interferometry.  (d) Time sequence of the triggers in the experimental setup
}
 \label{Procedure}
\end{figure*}
To overcome these limitations, we combine the LIAD launching technique, which provides liquid-free ejection of particles, with temporal control of the trapping potential, which overcomes limitations imposed by the absence of damping. We experimentally demonstrate this technique by loading nanospheres into a Paul trap. A schematic of the LIAD procedure is shown in Fig.~\ref{Procedure}a. A dry source of nanoparticles is prepared by pipetting a sonicated ethanol solution ($\sim 10^{10}$ particles/ml) containing silica nanospheres of radius $R = \SI[separate-uncertainty = true]{150+-20}{\nano\meter}$ and density of $\SI{2.0}{\gram\per\cubic\centi\meter}$ (Bangs Laboratories, Inc.) on a piece of $\SI{250}{\micro\meter}$ thick aluminum foil. After drying, the foil is mounted in the vacuum chamber at a distance $L = \SI{4.1}{\milli\meter}$ from the trap center, such that a normal to the foil is perpendicular to the axis of gravity, which is parallel to the trap axis. The chamber is then evacuated. The distance between opposite radio-frequency (RF) electrodes of the trap is $2r_0 = \SI{1.8}{\milli\meter}$, while the distance between the end-cap electrodes is $2z_0 = \SI{2.8}{\milli\meter}$.   Focusing a pulsed frequency-doubled Nd:YAG laser (\SI{5}{\nano\second} pulses of  \SI{3}{\milli\joule} at \SI{532}{\nano\meter}) down to \SI{200}{\micro\meter} on the back side of the foil, we create an acoustic wave that launches the particles from the front side via acoustic desorption~\cite{Millen2016}.  Figure~\ref{Procedure}b shows a scanning-electron micrograph of a nanosphere-coated foil before and after launching the deposited nanoparticles with ten laser pulses. The desorbed particles fly across the trap center, where they are illuminated with a \SI{635}{\nano\meter} laser and imaged using a CMOS camera (Fig.~\ref{Procedure}c). Particles with a charge-to-mass ratio lying in the stability zone of the Paul trap~\cite{Ghosh1995} can be trapped. With typical RF drive amplitudes of \SI{700}{\volt} and frequencies of \SI{7}{\kilo\hertz} applied to our trap electrodes, the upper limit of the charge-to-mass ratio is \SI{1}{\coulomb\per\kilogram}, which, in the case of \SI{300}{\nano\meter} diameter silica particles, corresponds to $\sim200$ elementary charges.

The LIAD method in low vacuum (0.1 mbar) has been tested for a wide range of particle sizes (100 nm $-$ 20 $\mu$m in diameter). To extend the procedure to high vacuum, we introduce temporal control of the trap potential. A similar procedure is used in experiments with trapped highly-charged atomic ions: Ionization typically occurs outside the trap volume, after which the ions are injected into the trap through holes in the end-cap electrodes, and control of DC electrode voltages enables ion capture~\cite{Schmoger2015}. In our setup, particles are injected along trajectories perpendicular to the end-cap electrodes (Fig.~\ref{Procedure}a and $xy$\nobreakdash-plane in Fig.~\ref{Procedure}c), and so we manipulate the trap's RF field instead of the DC end-cap voltages.
Fig.~\ref{Procedure}d shows a sketch of the time-delay sequence used. A first trigger signal activates the laser pulse, followed after a time $\Delta t$ by a second trigger signal that switches on the Paul trap. As a result, only particles with $xy$ speed component $v \approx L / \Delta t$ will be in the vicinity of the trap center when the trap is activated. If the sum of a particle's kinetic and potential energy is lower than the potential depth of the trap, the particle will be captured. With this procedure we to launch and capture particles at pressures as low as $\SI{4e-7}{\milli\bar}$, only limited by the pressure we can currently achieve with our setup.

The camera imaging allows us to track a trapped particle's center of mass (CoM) motion up to amplitudes of \SI{3}{\milli\meter}, but only up to frequencies of \SI{50}{\hertz}.  For motional resolution up to \SI{10}{\mega\hertz} frequencies along three axes, we illuminate the trap center with a focused \SI{1064}{\nano\meter} laser beam and use a detection system based on back-focal-plane (BFP) interferometry~\cite{Gittes1998} (Fig.~\ref{Procedure}c).
BFP interferometry accesses a more limited amplitude range than our camera, providing detection of only micron-sized orbits.

\begin{figure}
\centering
 \subfigure{\includegraphics[width=0.45\columnwidth]{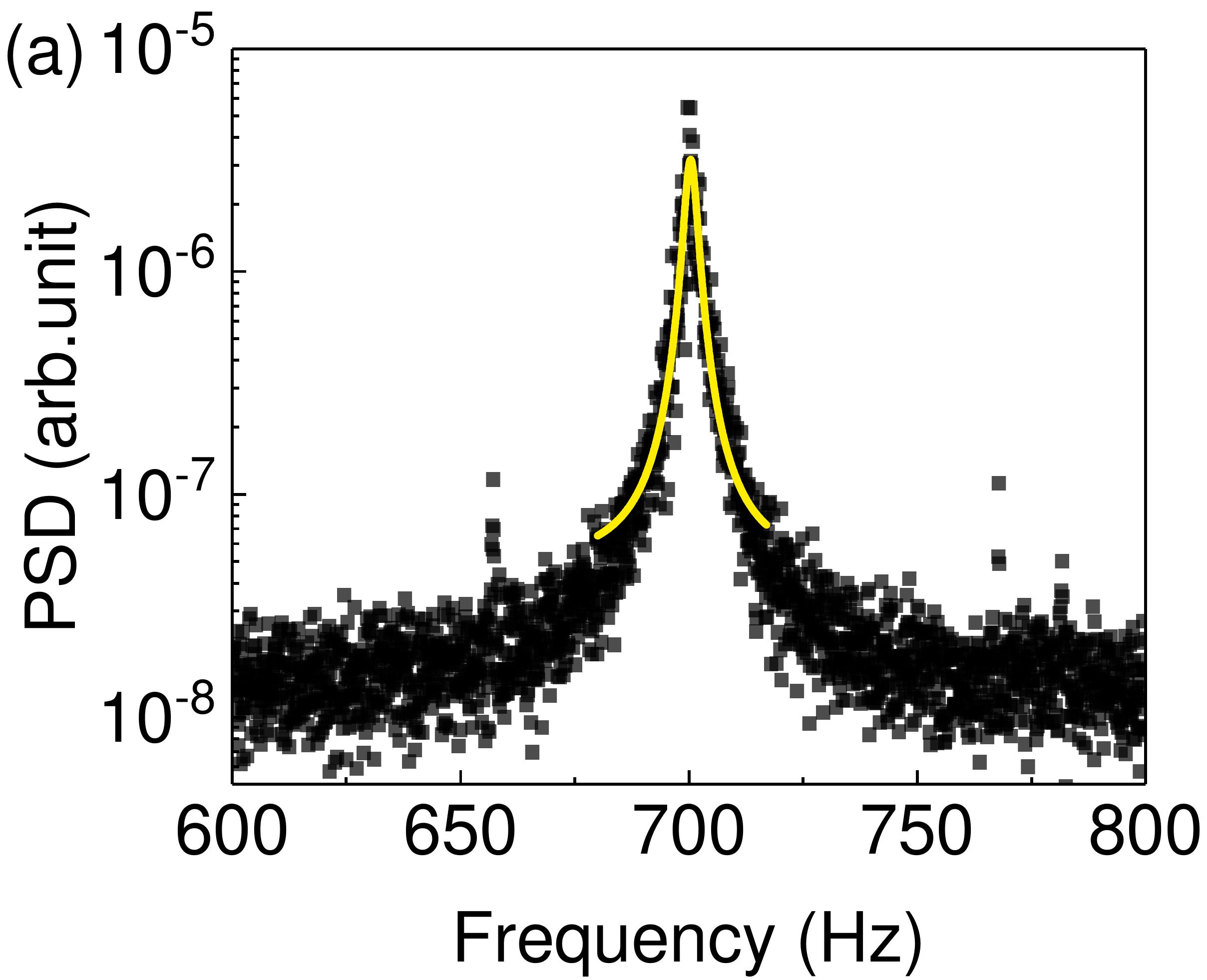}}
 \subfigure{\includegraphics[width=0.45\columnwidth]{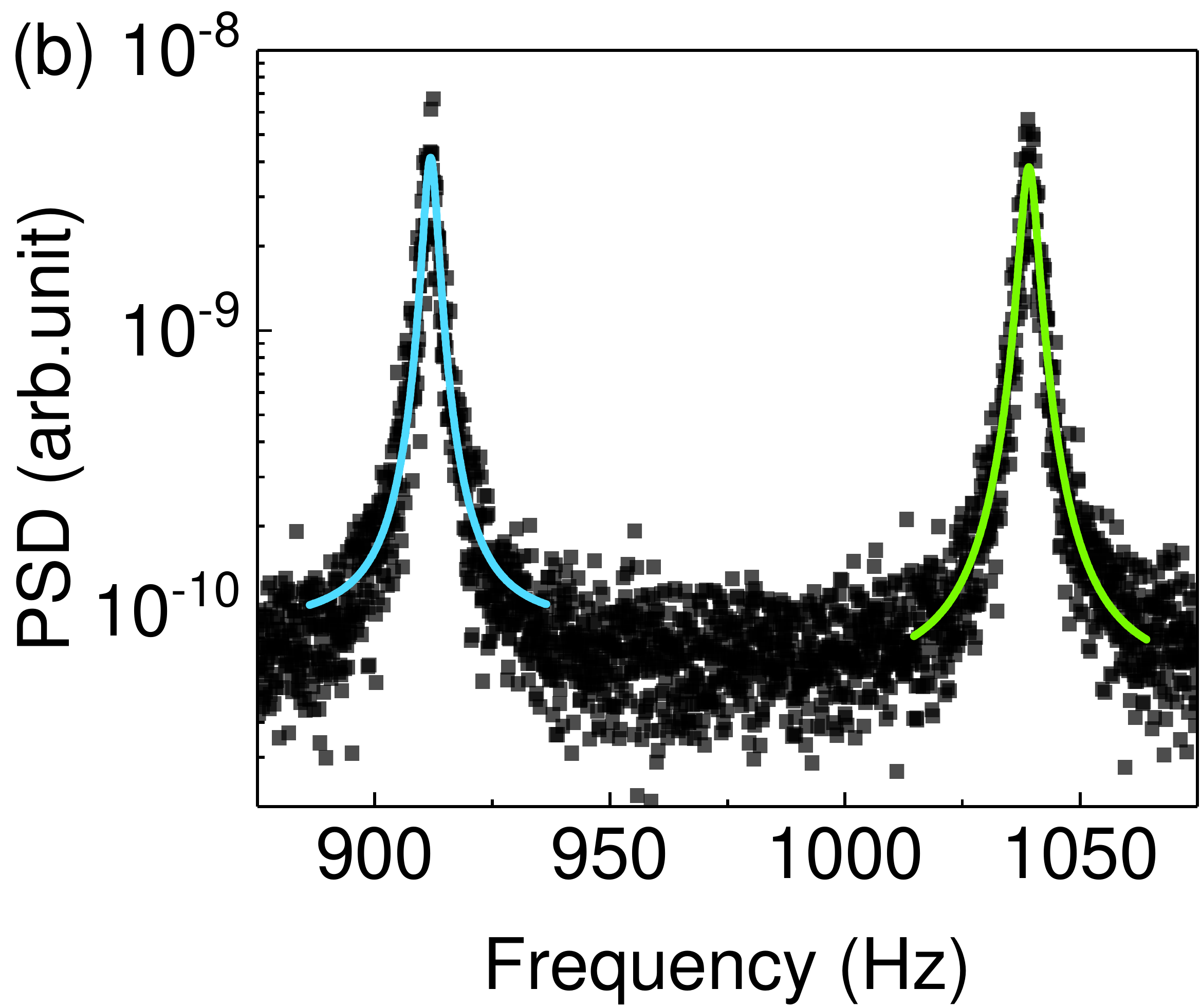}}
 \caption{Power spectral density of the CoM motion of a nanoparticle (a) along the trap axis and (b) in the radial plane of the trap. Measurements are taken at \SI{7e-3}{\milli\bar}. Solid lines represent fitting the experimental data with Eq.~\ref{ReponseFunction}.}
 \label{PSD}
\end{figure}

The equation for the secular motion of a charged particle levitated in a Paul trap~\cite{Ghosh1995} is $\ddot{x} + 2\gamma \dot{x} + \omega_0^2 x = F(t)/m$, where $\gamma = 3 \pi R \eta / m$ is the Stokes damping coefficient, $R$ is the radius of the particle, $\eta$ is the pressure-dependent dynamic viscosity of the gas, $\omega_0 = 2 \pi f_0$ is the resonance frequency, $m$ is the particle mass, and $F(t)$ is a driving force. Analyzing the thermal CoM motion in the frequency domain allows us to extract $\gamma$ and $\omega_0$ by fitting the experimental data with the single-sided power spectral density function~\cite{LiKheifetsRaizen2011,Berg-Sorensen2004}
\begin{equation}\label{ReponseFunction}
  g(\omega) = \frac{4 \gamma k_B T_0}{\pi m}\frac{A}{\left(\omega^2 - \omega_0^2\right)^2 + 4\gamma^2 \omega^2},
\end{equation}
where $k_B$ is the Boltzmann constant, $T_0$ is the bath temperature, and $A$ is a scaling constant. Examples of such fits along the three axes of particle motion are shown in Fig.~\ref{PSD}.
If the parameters $\omega_0, r_0, z_0$, and $m$ are known, we can estimate the trap depth.
For the data shown in Fig.~\ref{PSD}, the depth is $\SI{3.2}{\kilo\electronvolt}$ ($\SI{3.7e7}{\kelvin}$) along the trap axis and $\SI{2.8}{\kilo\electronvolt}$ ($\SI{3.2e7}{\kelvin}$) in the radial plane.

Note that the resonance frequency and thus the trap depth depend on the particle charge.
As a consequence, the resonance frequencies allow us to estimate the number of charges on the levitated particle. For the radial direction, the frequency is given by~\cite{Ghosh1995}
\begin{equation}\label{PaulTrapSResonance}
  \omega_0 = \frac{\Omega_{RF}}{2}\sqrt{-\epsilon_{DC} + \frac{\delta^2_{RF}}{2}},
\end{equation}
where
\begin{eqnarray*}
  \epsilon_{DC} = \frac{8q}{m\Omega_{RF}^2} k_{DC} \frac{V_{DC} / 2}{2 z_0^2};~ \delta_{RF} = \frac{4q}{m\Omega_{RF}^2} k_{RF} \frac{V_{RF} / 2}{r_0^2};
\end{eqnarray*}
$\Omega_{RF} = 2\pi \times \SI{7.7}{\kilo\hertz}$ is the RF drive frequency; $q$ is the particle charge; $k_{DC} = 0.38$ is a geometric factor accounting for the non-ideal shape of the end-cap electrodes, obtained from numerical simulations; $k_{RF} = 0.93$ is an equivalent geometric factor accounting for the RF electrodes; $V_{DC} = \SI{130}{\volt}$ is the voltage on the end-cap electrodes; and $V_{RF} = \SI{830}{\volt}$ is the RF drive amplitude. By substituting all parameters into Eq.~\ref{PaulTrapSResonance}, we find that the charge is between 85 and 95 elementary charges. The two value of charge come from the non-degenerate radial secular frequencies, which can be explained by asymmetries in the trap geometry unaccounted for by Eq.~\ref{PaulTrapSResonance}.

\begin{figure}
\centering
 \subfigure{\includegraphics[width=0.45\columnwidth]{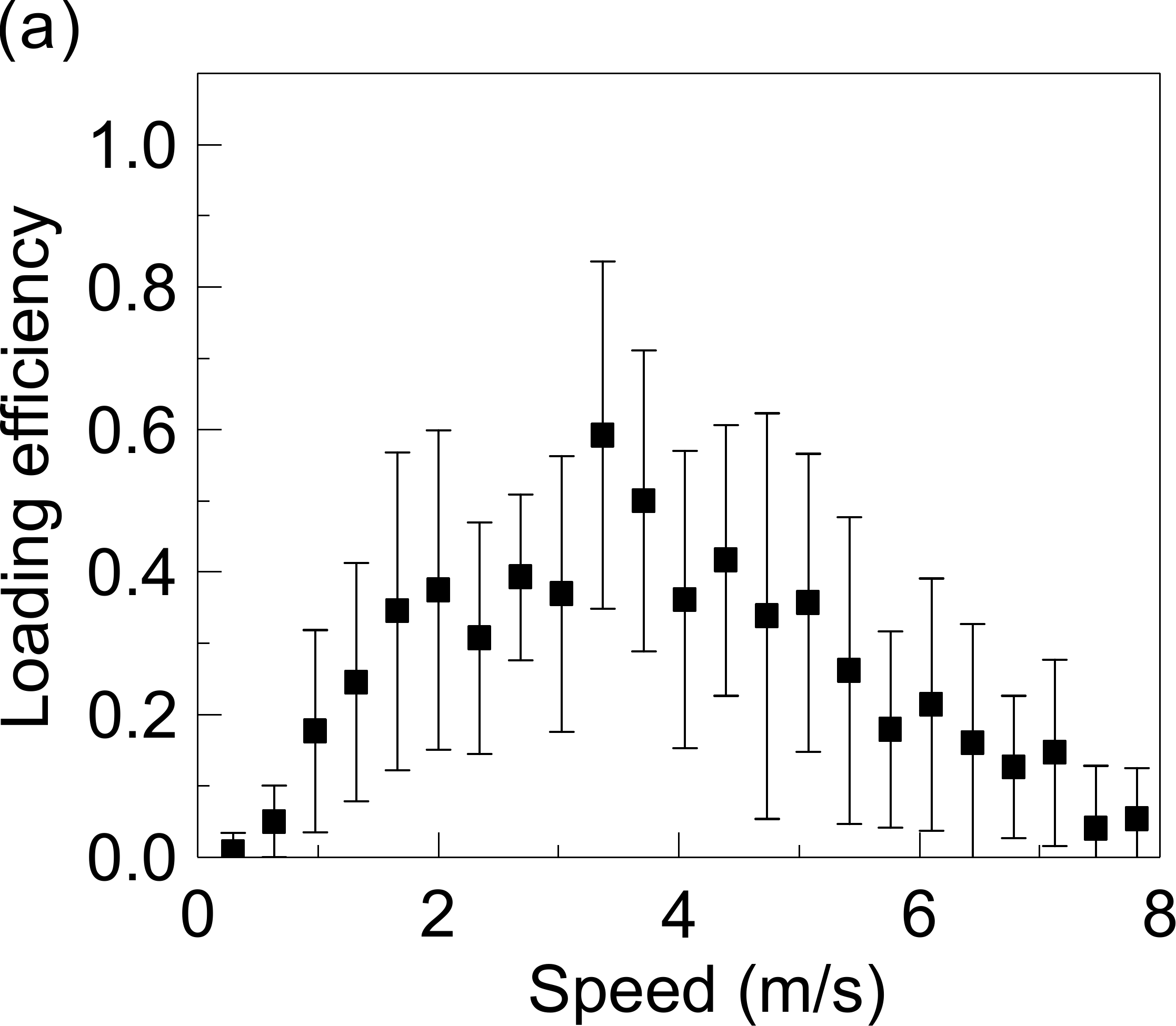}}
 \subfigure{\includegraphics[width=0.45\columnwidth]{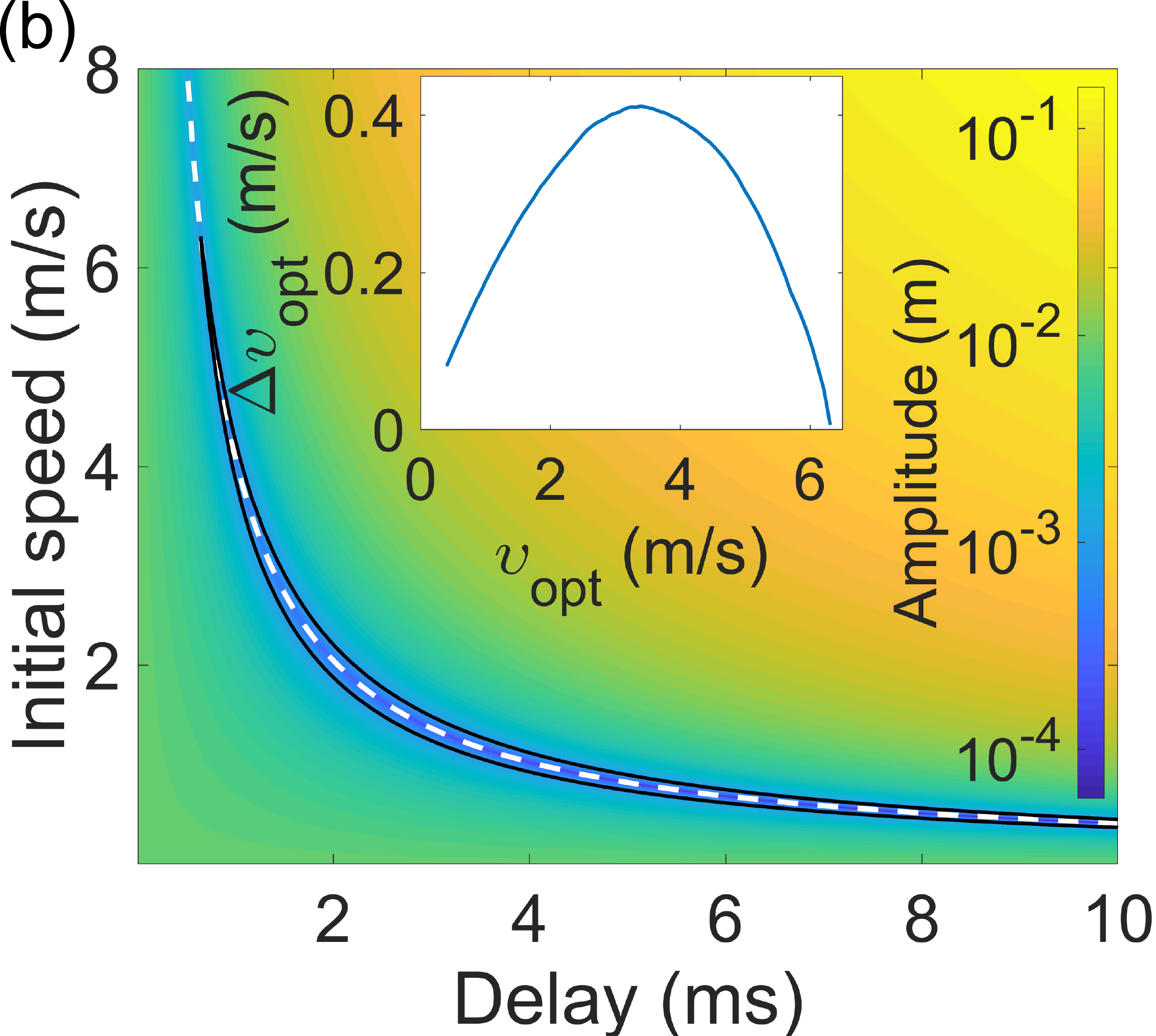}}
 \caption{Loading efficiency for different $xy$ speeds and delay times $\Delta t$. The loading efficiency is defined as the ratio of the number of successful trapping events to the number of total attempts. (a) Measured efficiency for different optimum particle speeds ($L / \Delta t$). (b) Simulated amplitude of the particle motion as a function of the delay time and initial speed after the trap is switched on. The white dashed curve follows the minima of the oscillation amplitude. The solid black lines delimit the region of the parameters that give an oscillation amplitude smaller than the trap size. Inset: region $\Delta v_{\text{opt}}$ around the optimum speed ($v_{\text{opt}}$) at which trapping is possible.}
 \label{Efficiency_speed}
\end{figure}

We now turn to the issue of particle loading efficiency.
Once a particle is launched from the foil, the probability that it will be captured depends on many parameters, including potential depth, the particle's initial velocity, and the time delay $\Delta t$ between the laser pulse and the trigger of the RF drive. The last two parameters are closely connected since $\Delta t$ determines the $xy$ component of the speed of the particles that will be inside the trapping region at the moment that the trap is turned on. In order to find the optimum trigger delay, we performed measurements of the loading efficiency for different values of $\Delta t$. We define loading efficiency as the ratio of the number of successful trapping events, which includes trapping of multiple particles, to the number of total attempts. The results are shown in Fig.~\ref{Efficiency_speed}a. The delay time $\Delta t$ is converted to particle speed: $v = L/\Delta t$.  Here, $v$ corresponds to the speed of a particle that is at the trap minimum when the trap is turned on. We find that a maximum trapping efficiency of \SI[separate-uncertainty = true]{0.6+-0.25}{} is achieved for particles with speeds of $\SI{3.4}{\meter\per\second}$, which corresponds to a delay time of \SI{1.2}{\milli\second}.
This speed corresponds to a kinetic energy of \SI{\sim 1}{\kilo\electronvolt}, well below the trap depth calculated for a particle with typical charge.

To understand the dependence of the trapping efficiency on the delay time and initial speed of the particles, we numerically simulated their trajectories in the plane orthogonal to the trap axis (the $xy$\nobreakdash-plane in Fig.~\ref{Procedure}c) during the loading process. Since the energy dissipated during the time of flight is negligible at the pressures we study (\SI{5e-3}{\milli\bar} for the data of Fig.~\ref{Efficiency_speed}a) compared to the initial kinetic energy, we neglected gas damping. 
Fig.~\ref{Efficiency_speed}b shows the amplitude of the particle's simulated oscillations as a function of the delay time $\Delta t$ and the initial launching speed. The minimum of the oscillation amplitude follows the relation $\Delta t = L/v$ (white dashed line), which corresponds to the trap being switched on when the particle is at the trap center. For each $\Delta t$, we find an optimum speed $v_{opt} = L / \Delta t$ and a speed range within which the particle's oscillations are smaller than the trap size, which indicates that the particle will be captured.  This region is delimited by solid black lines. The inset shows the dependence of the width of this region, $\Delta v_{opt}$, on $v_{opt}$, which peaks at $v_{opt}^{max}=\SI{3.4}{\meter\per\second}$. If all speeds that we consider are equally probable, the highest loading efficiency would be observed at $v_{opt}^{max}$, which agrees with the experimental data. This agreement suggests that the speed distribution in the vicinity of our working point is indeed close to uniform.

\begin{figure}
\centering
\subfigure{\includegraphics[width=0.45\columnwidth]{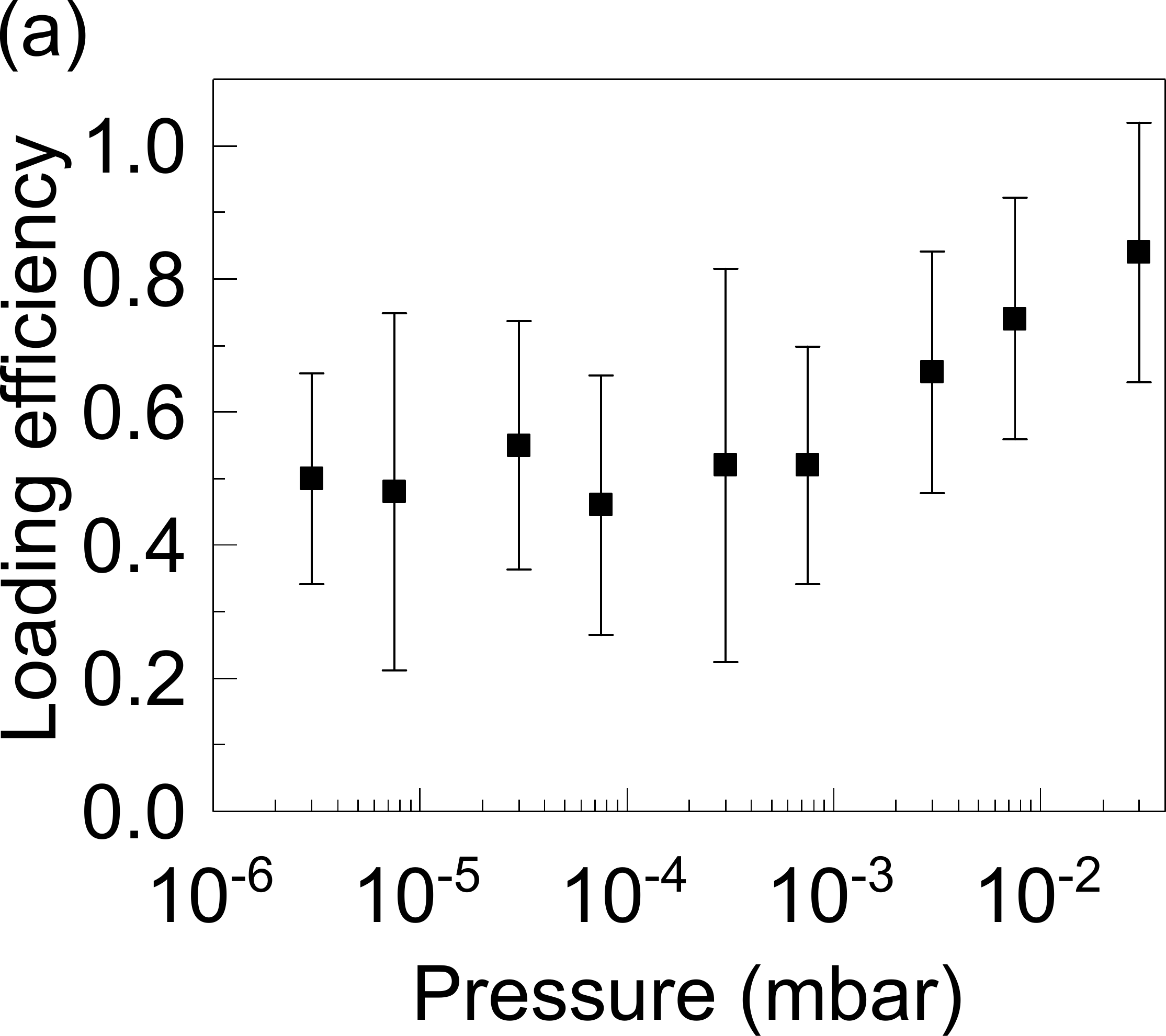}}
\subfigure{\includegraphics[width=0.45\columnwidth]{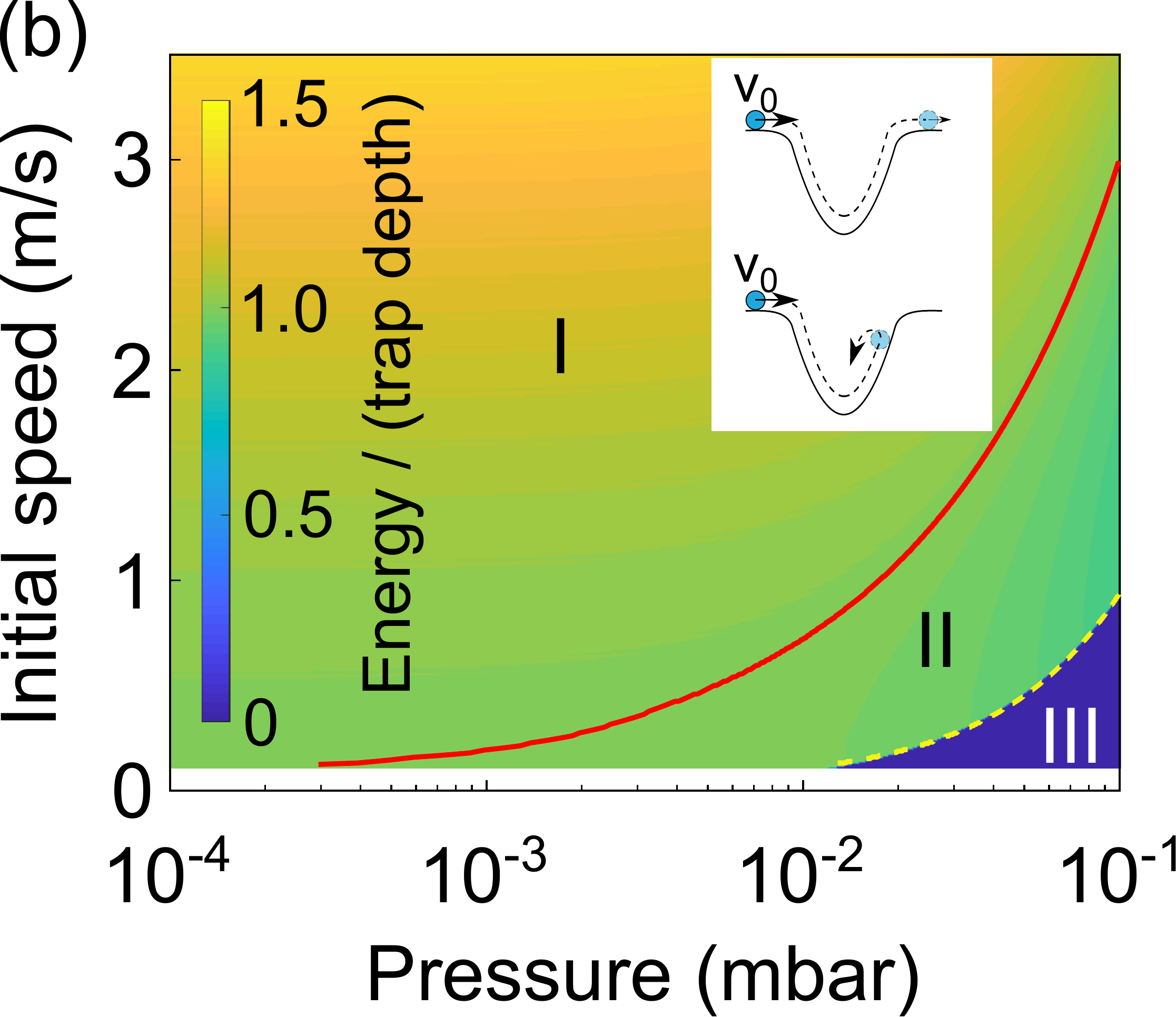}}
 \caption{Trapping efficiency at different pressures. A detailed description of the figure is given in the main text. (a) Measured trapping efficiency as a function of pressure. (b) Calculated energy of particles propagated towards and through the Paul trap potential as a function of pressure and initial speed. For the parameter spaces I and II, the energy is calculated at the far right point of the trajectory (inset). For the region III, the energy is set to zero.}
 \label{Efficiency_pressure}
\end{figure}

To investigate the pressure dependence of our loading method, we fixed $\Delta t$ to the optimum value of \SI{1.2}{\milli\second} and measured the trapping efficiency at different pressures. Fig.~\ref{Efficiency_pressure}a shows that the trapping efficiency for the optimal delay does not depend on the pressure in the range \SIrange[range-units = single]{e-6}{e-3}{\milli\bar}, while at pressures \SI{e-2}{\milli\bar} and above, an increase in the efficiency is observed. This effect can be explained by the increasing influence of the buffer gas on the particle motion. While at low pressures, most captured particles were already inside the trapping region at the moment the trap was switched on, at higher pressures, additional trapping events occur due to slow particles that arrive later. For a better understanding of this mechanism, we simulated propagation of particles towards and through the potential of the Paul trap. To separate the effects of pressure from those of temporal control, we considered the trap to be always on. The numerical simulations reveal three types of particle motion. First, if the pressure and initial speed lie in region I of Fig.~\ref{Efficiency_pressure}b, the particle crosses the trap without being captured. Second, if these parameters lie in region III, the particle is stopped by friction before it reaches the trap. Third, and most relevant for us, if the parameters lie in region II, the particle enters the trap and dissipates enough energy by friction that it remains trapped. From Fig.~\ref{Efficiency_pressure}b, we see that the range of speeds in region II grows as one moves towards higher pressures, and particles with higher speeds contribute to the trapping events. Note that we did not acquire data at pressures below \SI{3e-6}{\milli\bar} because simulations support our observations that the loading efficiency is independent of pressure below \SI{e-3}{\milli\bar}, and because accumulating statistics on loading efficiency becomes increasingly time-intensive with decreasing pressure.
\begin{figure}[h!]
 \includegraphics[width=0.9\columnwidth]{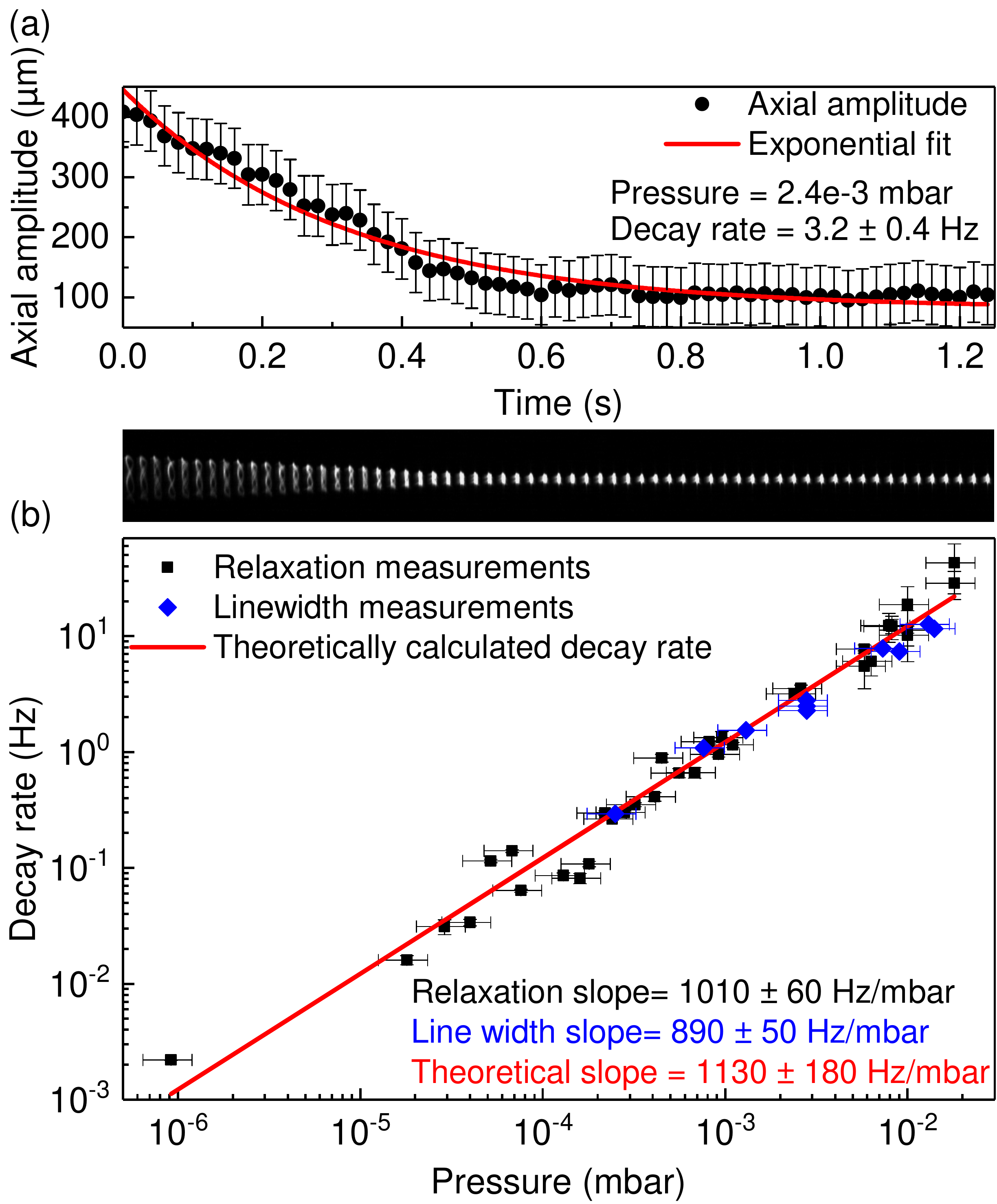}
 \caption{Thermalization of loaded particles to the room-temperature background gas. (a). An example of a decay of particle's oscillation amplitude (top) extracted from a recorded video. Corresponding frame sequence (bottom). (b) Measured amplitude decay rate at different pressures.}
 \label{Decay_rate}
\end{figure}

Besides pressure and particle speed, another factor that affects the loading efficiency is the ability of a particle to dissipate its high initial energy, which can be studied by measuring the relaxation of the CoM motion of the particle after it is captured. Such measurements allow us to ascertain whether the particle thermalises to the background gas or other heating and cooling sources are present, such as heating due to the Paul trap itself.  We launched particles over a range of pressures and recorded their trajectories as videos. From these, we extracted the amplitude of vertical ($z$-axis) motion, which is plotted versus time in Fig.~\ref{Decay_rate}a. By fitting the decay with the exponential function $z = z_{off} + z_0 e^{-\gamma t}$, we extract the damping coefficient $\gamma$. We find that the damping coefficient depends linearly on pressure with a slope of \SI[separate-uncertainty = true]{1010+-60}{\hertz\per\milli\bar}. This value agrees with the value \SI[separate-uncertainty = true]{890+-50}{\hertz\per\milli\bar} extracted from resonance curves of the particle's thermally driven motion at different pressures, and with the slope of \SI[separate-uncertainty = true]{1130+-150}{\hertz\per\milli\bar}, obtained from the theoretical prediction~\cite{Beresnev1990,JainGieselerMoritzEtAl2016}
$\gamma = 7.9 R^2 P / m \upsilon_{gas}$,
where $P$ is the gas pressure, and $\upsilon_{gas}$ is the root-mean-square velocity of the gas molecules (Fig.~\ref{Decay_rate}b).
This agreement suggests that the relaxation of the particle's CoM motion is dominated by background gas cooling, and the trapping mechanism induces no additional reheating. Extrapolating the damping rate to lower pressures, we estimate the time needed to thermalize with the background gas to be 5.5 years at $\SI{e-11}{\milli\bar}$. Therefore, an additional cooling mechanism would be required to use the method in UHV. This cooling mechanism could be provided by a cold noble gas supplied to the trap center through a retractable tube, parametric feedback damping~\cite{nagornykh2015cooling} or cold damping~\cite{BushevCooling}.

To conclude, we have demonstrated a method that allowed us to load nanoparticles directly in a Paul trap at high vacuum conditions. The minimum pressure at which we observed loading events was \SI{4e-7}{\milli\bar}, limited by the pressure our current vacuum chamber can reach. Pressures down to UHV levels are expected in the current chamber following bakeout and the addition of a combined ion-getter pump. The chamber can be baked out with the dry source of the particles already mounted inside. Since the dry source consists only of aluminium foil and silica nanoparticles, the source does not limit the bakeout procedure.
Our measurements both allow us to optimize experimental parameters and demonstrate the high efficiency of this technique: The loading process takes less than a second, orders of magnitude faster than in other currently available setups~\cite{Kane2010,LiKheifetsRaizen2011,gieseler_subkelvin_2012,KieselBlaserDelicEtAl2013, MillenFonsecaMavrogordatosEtAl2015,MestresBerthelotSpasenovicEtAl2015,Vovrosh2017}.
Our method is promising for UHV when combined with an additional damping mechanism to cool the particle's CoM.
In turn, a UHV setting is expected to enable novel experiments for levitated nanoparticles in a quantum regime, including the preparation of nonclassical motional states, ultrasensitive detectors~\cite{ChangRegalPappEtAl2010,Romero-Isart2011a}, and tests of quantum gravity\cite{Marletto2017, Bose2017}.

\begin{acknowledgments}
 We thank S. Kuhn for valuable advice and insightful discussions.  This work was supported by the Austrian Science Fund (FWF) Project Y951-N36.
\end{acknowledgments}


%
%

%


\bibliography{Direct_loading_paper}

\end{document}